\documentclass[floatfix, preprintnumbers, amsmath, amssymb, twocolumn, 10pt]{revtex4}
\usepackage{bm}
\begin{document}
\date{\today}

\newcommand{\eq}[2]{\begin{equation}\label{#1}{#2}\end{equation}}

\title{Exponential Inflation With $\rho = +p$}

\author{Subodh P. Patil $^{1,2)}$} \email[email: ]{patil@het.brown.edu}

\affiliation{1) Dept.of Physics, Brown University, 182 Hope Street,
Providence, R.I. 02912, U.S.A.} 
\affiliation{2) Dept.of Physics, McGill University, 3600 University St., Montr\'eal QC, H3A 2T8, Canada.}

\begin{abstract}
We introduce in this paper a new framework for obtaining a period of exponential inflation that is entirely driven by the quadratic kinetic energy of a scalar field. In contrast to recent attempts to realize scalar field inflation without potentials (such as k-inflation or ghost inflation), we find that it is possible to obtain exponential inflation, without invoking any higher derivative actions, or modifying gravity, and that unlike all previous approaches, we do not require a $\rho = -p$ phase in order to realize inflation. The inflaton in our proposed framework is a scalar field with a quadratic kinetic energy term, but with the `wrong sign'. We take the perspective that this is due to some temporary instability in our system at high energies, and provide physical examples of situations where a modulus field might temprorarily exhibit such behaviour. The deflation of extra dimensions is a neccesary feature of our framework. However, unlike in previous attempts at Kaluza-Klein inflation, it is possible to obtain exponential (as opposed to power law or pole) inflation. We provide several indications of how one can gracefully exit from this type of inflation. 
\end{abstract}

\maketitle

\section{Introduction}

Inflation \cite{guth}\cite{linde1}\cite{linde2}, in spite of its spectacular successes in accounting for the initial conditions of the hot big bang, as well as the observed anisotropies of the cosmic microwave background radiation (CMBR), remains a paradigm without a solid theoretical realization. Indeed, the usual scalar field driven inflation that we take for granted is faced with several (quite severe) conceptual difficulties. One such difficulty is the trans-planckian problem, where one has to confront the fact that scales that are observable today were inflated up from sub Planck length scales, and as such should have to account for the new physics that is bound to effect such modes. From a particle physicist's point of view, an additional issue faced by inflation is the problem of generically having to usually fine tune the potential of the inflaton to one part in $10^{-12}$, which is closely related to the problem of trying to find a suitable candidate for the inflaton among known or conjectured models of particle, or stringy physics. Another concern for inflation includes the fact that in the case of single field inflation, the scalar field driving inflation must have initial values greater than the Planck scale \footnote{We wish to remind the reader that in general, introducing multi-field inflation, which generically avoids this problem neccesarily involves, in the absence of some other mechanism, introducing entropy perturbations \cite{rhb2}, which are not observed in present experiments.}. Also, in the absence of new physics, inflationary cosmology offers no solution to the singularity problem\cite{borde1}\cite{borde2} (see \cite{rhb1}\cite{rhb3} for a nice review of the points raised here). Clearly, although inflation as a paradigm has proved its worth in terms of accounting for observations, we still have a long way to go before we can claim that we have understood this aspect of early universe physics.

\par

However, one particular challenge facing standard scalar field driven inflation, is intimately tied up with one of {\it the} outstanding problems facing cosmology today: the cosmological constant problem \cite{sw}\cite{carroll}. The challenge can be stated as follows: why does the time independent part of the potential energy of a scalar field generate inflation, while the quantum vacuum energy, which is another source of constant potential energy (in the absence of a consistent mechanism which renders it to be vanishingly small at present energies), remains gravitationally inert? Clearly, if standard potential-driven scalar field inflation is responsible for the large scale structure and the initial conditions of our universe, then somehow the dynamics of gravity must render one form of potential energy gravitationally non-dynamical whilst responding to the other \cite{rhb3}. Thus one might have the concern that any future solution of the cosmological constant problem might neutralize the mechanism responsible for potential-driven inflation.

\par

It is the goal of this paper to present a mechanism in which one could obtain exponential inflation with a $\rho = +p$ fluid. Such a fluid will not have the time independent part of its energy momentum tensor to be proportional to the metric tensor, and hence will not look like a cosmological constant. At the very least, this presents us with a novel way in which one could realize inflation, and allows us more leeway in our search for the true inflaton. More importantly however, were we to realize inflation in this manner, then we will have severed the cosmological constant problem from the problem of finding a consistent realisation of the inflationary paradigm. We find that, in addition, one avoids having to fine tune our model specifically in the particle physics sector, as the usual place where such fine tuning occurs (potentials) does not feature anymore. Although we will not have completely rid ourselves of having to fine tune, it is conceptually quite appealing not to have to tune any potentials, as one expects that at unification scales, particle physics may not neccesarily be open to such tweaking. 

\par

In order to give us perspective on the results that are to follow, we offer a brief survey of the various attempts to realize inflation without potentials, and how these might differ from the mechanism we propose.

\par

The general philosophy of kinetically driven inflation is to to obviate the need for potentials, and to obtain inflation purely through the kinetic energy term of some scalar field, or solely through the dynamics of the gravitational sector of some theory. A potential advantage of this approach is that, from a phenomenological viewpoint, it is easier to find massless scalars in high energy theories such as string theory (i.e. moduli fields), than it is to find scalar fields with potentials that have the requisite properties to make them candidates for the standard `slow roll' inflaton. Two such approaches to kinetically driven inflation have been proposed of late: k-inflation \cite{k}, and ghost inflation \cite{nah1}, where the latter finds its motivation not in string theory, but in an attempt to construct a consistent infra-red modification of gravity \cite{nah2}. Both of these approaches rely upon higher derivative kinetic terms of some scalar field (some modulus field in string theory in the case of k-inflation, and a ghost condensate in the case of ghost inflation), to obtain a $\rho = -p$ equation of state which will generate a de Sitter phase of evolution. A consequence of this higher derivative dependence is that it will only be operative in very high energy regimes, which will have obvious observational consequences in terms of the gravitational wave background (which is predicted to be quite large), as well as non-gaussianities in the CMBR, due to the presence of higher derivative kinetic terms.

\par

In \cite{GV}, in the context of dilaton gravity, power-law inflation was obtained purely from the dynamics of the free dilaton, albeit in the `string frame' \footnote{Recall that in the low energy supergravity approximations to string theory, one has to conformally rescale the metric to go from the string frame, where the gravitational theory appears as a Jordan-Brans-Dicke type theory, to the Einstein frame. This usually entails a transition from an inflating type solution to a deflating one.}. In \cite{levin}, the idea (which has been around for some time \cite{ellis1}\cite{ellis2}\cite{kolb}) that one can use the deflation of extra dimensions to generate inflation in the non-compact directions, was used to yield pole inflation (where the Hubble factor diverges in finite time) \footnote{Related approaches include that of \cite{pollock}, where super-exponential inflation is obtained in higher derivative gravity, although potential like terms appear upon dimensional reduction, and inflation in $f(R)$ gravity (\cite{rhb2} and references therein), where again, potential like terms make their appearance in the frame in which inflation is obtained.}. Common to both these approaches is that one is doing away with matter sources altogether to generate inflation, the dynamics of spacetime being generated completely from gravity itself. Also common to these approaches, is that one does not obtain exponential inflation. In terms of reproducing observation, this is not such a desirable property, as one would have to tune this pole law, or power law behaviour to approximate an exponentially expanding epoch for long enough. This is because, in general, inflation that is either `faster' than or `slower' than exponential, will yield a tilt in the spectrum of metric fluctuations away from scale invariance.
\par
 
The inflationary framework we propose in this report involves the use of a scalar field, which might (in certain cases), be view as a modulus field, with a temporarily `wrong sign' kinetic term. Our perspective on this is that this negative kinetic energy is either indicative of an instability in the system, or is due to some temporary state of affairs in which the overall dynamics of the system remains well defined and bounded. In this sense, our mechanism can be viewed as loosely analogous to tachyon inflation. We furnish several examples where scalar degrees of freedom temporarily acquire negative kinetic energies, e.g. models where branes with negative tension propagate through a compact manifold, theories describing the motion of BPS monopoles, supergravity theories or higher derivative theories coupled to gravity and certain string field theoretical setups.


\par

The outline of the paper is as follows: we begin with the demonstration that a scalar field with a negative definite kinetic energy yields a period of exponential inflation of certain dimensions provided certain other dimensions deflate. We then discuss the physical sensibility of our framework, drawing upon the examples just mentioned to provide models of how this mechanism could be realised in the early universe. After we discuss some consequences of our results, and the issue of how one could gracefully exit from this type of inflation, we comment on the nature of this type of inflation as a dynamical compactification mechanism for extra dimensions. We then speculate on observational predictions of this framework.

\par

We wish to emphasize that this report is in the spirit of a preliminary investigation, as many of the more interesting consequences of this type of inflation will depend on its precise realization (we offer several candidate mechanisms further on), and shall occupy us in a future study. In the interests of berevity, we shall sometimes refer to the type of inflation that we present in what follows as `$w = 1$ inflation'.   

\par

Our conventions are as follows: we work in natural units ($\hbar = c = 1$), with the units of energy measured in electron volts (eV). Our universe is taken to have $D$ spatial dimensions (i.e. $D+1$ spacetime dimensions), with $G_D$ being the corresponding $D$-dimensional Newton's constant. We also use the `East Coast' convention for our metric tensor: $g_{\mu\nu} = (-,+,+,+,..,+)$.

\section{The Framework-- $w = 1$ Inflation}

\par We begin by considering Einstein gravity minimally coupled to a scalar field with the `wrong sign' for it's kinetic energy (whose utility will become clear shortly):  

\begin{equation}
\label{act}
S = \frac{1}{16\pi G_D} \int d^{D+1}x \sqrt{-g}R + \frac{1}{2}\int d^{D+1}x \sqrt{-g} \partial_\mu\phi\partial^\mu\phi.
\end{equation}

\noindent The energy momentum tensor of the scalar field in the above set up yields

\begin{equation}
\label{emt}
\rho = p = -\frac{\dot{\phi}^2}{2},
\end{equation}

\noindent where we draw attention to the fact that the equation of state parameter $w$ is identically

\eq{w}{w = 1.}

\noindent We now make the metric ansatz

\begin{equation}
\label{metric}
ds^2 = -dt^2 + \sum_{i=1}^{D} a_i^2(t)dx_i^2,
\end{equation}     

\noindent where $D$ is the total number of spatial dimensions, with $n~$ compact directions, and $d$ non-compact directions, both taken to be separately isotropic. The resulting Einstein equations ($G^\mu_\nu = 8\pi G_DT^\mu_\nu$) can be recast in the form:

\begin{equation}
\label{ee}
\ddot{a}_i + \dot{a}_i\Bigl(\sum_{j\neq i} \frac{\dot{a_j}}{{a_j}} \Bigr) = 8\pi G_Da_i\Bigr[p_i-\frac{1}{D-1}\sum_{j=1}^{D} p_j+\frac{1}{D-1}\rho \Bigl],
\end{equation}

\begin{equation}
\label{00}
\rho = \frac{1}{16\pi G_D}\bigl[(\sum_\mu H_\mu)^2 - \sum_\mu (H_\mu)^2 \bigr].
\end{equation}

\noindent Notice that in (\ref{ee}), the above form for the energy momentum tensor (\ref{emt}) drops out of the driving term. Hence, the diagonal Einstein equations for the spatial components become inistinguishable from that of the vacuum:

\begin{equation}
\label{s1}
\dot{H} + (dH + n\mathcal H)H = 0,
\end{equation}
\begin{equation}
\label{s2}
\dot{\mathcal H} + (dH + n\mathcal H)\mathcal H = 0,
\end{equation} 

\noindent where $H$ is the Hubble factor associated with the scale factor of the non-compact directions, $a(t)$, defined as $H = \dot a/a$, and $\mathcal H = \dot{b}/b$ is the Hubble factor of the compact directions with $b(t)$ the corresponding scale factor. Hence we see that if

\eq{vct}{dH + n\mathcal H = 0,} 

\noindent that is, if the volume of spacetime is constant, and $\dot H = \dot{\mathcal H} = \ddot\phi = 0$, we solve all of the Einstein equations, with inflation in the non-compact directions, provided the compact dimensions correspondingly deflate. However the $00$ equation (\ref{00}) reads

\begin{eqnarray}
\label{o0} -8\pi G_D \dot{\phi}^2 &=& -dH^2 - n\mathcal H^2,\\ \nonumber
&=& -d(1+ d/n)H^2.
\end{eqnarray}

\noindent Hence we see the neccesity to introduce wrong sign kinetic term-- the above would not be possible to satisfy with the right sign kinetic term. Additionally, we see that although as far as the space-space Einstein equations (\ref{s1})(\ref{s2}) are concerned, that although the scalar field background is no different from that of the vacuum, the $00$ equation would not be satisfied with our inflationary ansatz in a purely vacuum setting. In \cite{levin} an inflationary setting was studied, where extra dimensions deflate whilst non-compact directions inflate in just such a vacuum setting. However, it was neccesary in this case to introduce time derivatives for the Hubble factors, with the effect that for greater than one extra dimension, pole inflation (where the scale factor $a$ diverges in finite time) was obtained. Here we see that by admitting a scalar field with a negative definite kinetic energy, one obtains exponential inflation with ease, the significance of which we shall discuss further on.

\par

Note that this ansatz is consistent with the equation of motion for the scalar field (= covariant conservation of energy-momentum tensor):

\begin{equation}
\ddot{\phi} + (dH + n\mathcal H)\dot\phi = 0 \to \ddot{\phi} = 0,
\end{equation}  

\noindent so that $\dot{\phi}$, $H$ and $\mathcal H$ are all constants. We emphasize here that we now have inflation {\it without} potentials. In fact, potentials would derail this setup-- we cannot simultaneously solve all of the Einstein equations and the equation of motion for the scalar field to obtain inflation for the non-compact directions with a non-zero kinetic energy and a non-zero potential for the scalar field. This is the origin of the `slow roll' approximation in standard inflation where, by setting the derivatives of the scalar field in question to be small enough, one could obtain an isotropically inflating solution. In this sense, our proposed mechanism can be considered opposite or complimentary to the standard potential-driven inflation. Before we get to the questions that are immediately begging, such as how one might gracefully exit and re-heat from this type of inflation, in addition to which if any of the observational predictions made within this framework (which will in general, be model dependent) differ from standard potential-driven inflation, we turn to the important question of the `wrong sign' kinetic energy. In the process, we shall also uncover several candidate models which might realize this framework.

\section{Scalar Fields With Temporarily `Wrong Sign' Kinetic Terms}

Although scalar fields with the `wrong sign' kinetic energy term have been widely used to model matter for which the equation of state parameter $w$ can be less than $-1$ (see \cite{CT} and references therein), the physical sense of using such a field with a Hamiltonian unbounded from below remains murky. In particular, it was shown in \cite{cline} that any such phantom field which survives at low energies (it was the intended use of such fields to explain the current acceleration of the universe), is phenomenologically inconsistent. However, there is a greater difficulty in introducing such matter fields, in that any form of matter which violates the dominant energy condition of general relativity (with $w \leq -1$) leads to a catastrophic instability of the vacuum. However, in order for a phantom field to have an equation of state parameter to be less than $-1$, one must neccesarily involve a potential for the field. We have already shown in (\ref{emt}) and (\ref{w}) that through introducing a scalar field with the wrong sign kinetic energy, {\it without potentials}, one finds that the equation of state parameter is precisely $1$, and hence does not cause immediate alarm in this respect. 

\par

One nevertheless has to worry about the fact that one is dealing with a Hamiltonian which is potentially unbounded from below. However, we presently wish to furnish several physical examples where one could be dealing with a system which admits a scalar degree of freedom that acquires a negative definite kinetic energy only temporarily, and as such, is dynamically stable. A few examples, that of (unstable) negative tension branes propagating in a compact manifold, and embedding our inflaton in some higher derivative theory, will also have the overall Hamiltonian of the system perenially bounded from below. We do not claim to have exhausted all of the possible realizations of this framework, rather, our goal is to motivate this framework through specific examples, and in doing so, to offer another avenue in which one can obtain inflation in stringy or other high energy models of the early universe. 

\subsection{BPS monopoles, sigma models and moduli space metrics}

Consider the following action describing some non-linear sigma model,

\eq{nlsm}{S = -\frac{1}{2\kappa}\int d^{D+1}x \sqrt{-g} G_{ij}(\phi)\partial_\mu \phi^i\partial_\nu\phi^j g^{\mu\nu},}

\noindent where $\kappa$ is some appropriate dimensionful parameter, and $G_{ij}(\phi)$ is the sigma model metric. It is usually the case that the metric $G_{ij}$ is positive definite, but consider the following situation: let us assume that over a finite region in field space, the metric acquires a negative eigenvalue,

\eq{neg}{det[G_{ij}(\phi) - s\delta_{ij}] = 0~;~ s < 0~,~ \phi~ \epsilon~ U,}

\noindent where $U$ is some finite, bounded region in field space. Let us also take for simplicity this bounded region to occupy some region around the origin in field space. In this case one finds that whenever the field vector enters this region, it acquires a negative definite kinetic energy. Assuming that there are interactions with some external source that cause the fields to acquire more and more negative kinetic energy, we will find that the field variables move faster and faster through field space in a runaway manner. The net effect of this is that the more negative energy the field acquires, the quicker it will exit the region where it has a negative definite kinetic energy. Hence in a dynamical sense this system is stable; the apparent instability is only a temporary feature since the field avoids this region in field space for all later times if it were in such a region initially. 

\par

Although seemingly far-fetched, the above example is qualitatively representative of the sigma model that results when one considers the motion of $n$ BPS monopoles (recall that BPS objects excert no forces upon each other, hence the analogy above is accurate in that no bound states are liable to form). It was shown in \cite{gibbons} that such a system can be described in terms of the Lagrangian, the quadratic part of which can be written as

\eq{bps}{L = \frac{1}{2}g_{ab}\dot{x}^i_a\dot{x}^i_b~;~ 1 \leq a \leq n,}

\noindent where the moduli space metric is given by:

\begin{eqnarray}
g_{aa} &=& 1 - \sum_{b\neq a} \frac{1}{|\vec x_a - \vec x_b|},\\
g_{ab} &=& \frac{1}{|\vec x_a - \vec x_b|}~;~ a \neq b,
\end{eqnarray}

\noindent which clearly exhibits the sort of behaviour indicated above. In this case, when enough of the monopoles get close together, the metric flips sign. As more energy gets dumped into the motion of these monopoles, the quicker they separate again, and the positivity of the metric is restored. Although there may exist other such similar physically realised sigma model metrics, we turn our attention to other situations where fields might temporarily exhibit such instabilities. Our next example is motivated by recent ideas in string cosmology \cite{st1}-\cite{turok}.

\subsection{Negative tension branes -- why instability can be a good thing}

Negative tension branes often arise in an orbifold setting, out of neccesity rather than through any real desire to have them around. For instance, in the context of 11 dimensional heterotic M-theory \cite{witten}, where 6 dimensions are compactified on a Calabi-Yau 3-fold and one dimension is compactified on an orbifold with two orbifold fixed planes and an internal brane, the simplest model one is allowed to look at (having set all the fields to zero for which it is consistent to do so) is the following:

\begin{eqnarray}
\label{hmt}\nonumber
S &=& \frac{M^3_5}{2}\int_{\mathcal M} d^5x \sqrt{-g}\Bigl(R - \frac{1}{2}\partial_a \Phi\partial^a\Phi - \frac{3}{2}\frac{e^{2\Phi}}{5!}F^2\Bigr)\\ &-& \sum_{i = 1}^{3} 3\alpha_iM^3_5\int_{\mathcal M_4^{(i)}} d^4\xi_i \sqrt{-h_i}e^{-\Phi}\\ \nonumber &-& \sum_{i = 1}^{3} \frac{3\alpha_iM^3_5}{4!}\int_{\mathcal M_4^{(i)}} d^4\xi_i\epsilon^{\mu\nu\lambda\beta}A_{abcd}\partial_\mu X_i^a\partial_\nu X_i^b\partial_\lambda X_i^c \partial_\beta X_i^d,
\end{eqnarray}

\noindent where we refer the reader to \cite{tseytlin} if the meaning of anything in the above is unclear. We note that the brane tensions (given by $\alpha_i$), have to satisfy through the equations of motion, the sum rule condition,

\eq{sr}{\sum_i \alpha_i = 0.}

\noindent Hence negative tension branes are a neccesary feature of this model. In fact, negative tension branes turn out to be generic features of gravitational theories compactified on orbifolds. In \cite{kogan2} it was shown that the radion modulus corresponding to a negative tension brane propagating through a compact orbifold has a `wrong sign' kinetic energy term associated with it. In \cite{turok}, a similar observation is made in a different setting (see also \cite{kogan1} and \cite{dvali} for further discussion). 

\par

The perspective that we wish to emphasize on the use of negative tension branes in cosmology is that unless they are describing orbifold fixed planes, they are intrinsically unstable objects. Hence, they are prone to decay, subject to the specifics of the theory in which these branes appear. This instability suits our purposes quite well, in that we have a scalar degree of freedom which temporarily has a negaitve definite kinetic energy. The eventual decay of this degree of freedom means that overall, one has a Hamiltonian which is bounded from below; the wrong sign for the kinetic energy manifests itself temporarily, as we have chosen to introduce an intrinsically unstable object into our framework, which upon decay results in a system with a well defined ground state.

\subsection{Examples from string theory and supergravity}

Scalar fields with wrong sign kinetic energy terms often appear in supergravity theories \cite{nilles}, especially when one is considering non-unitary representations of certain superalgebras \cite{hull}. However, it could be that the theories in which these phantom fields appear are related by some duality transformation to a more standard supergravity theory \cite{hull}, which remains free of pathologies. Their interpretation and uses are not clear to us at this time, so we will avoid any reference to them. However, in \cite{arefeva} (and references therein), it is shown that `phantom' like fields appear at high energies in a string field theoretical setting. Consider a kinetic term for a scalar field of the standard form

\eq{kt}{S = \int d^{D+1}x~ \kappa^2\phi\square\phi~;~\kappa^2 < 1,}

\noindent where due to non-local factors in the interaction terms for the various string modes, it is desirable to make a field redefinition which causes the kinetic term in the above to transform into

\eq{ktt}{\kappa^2\phi\square\phi ~\to~ \phi(\kappa^2\square + 1)e^{-\square}\phi~~;~~\kappa^2 < 1.}

\noindent Hence, were we to make a `local' approximation and keep only terms up to second derivatives, we find that we acquire a wrong sign for the kinetic energy, while the full high energy theory is well defined \cite{arefeva2}. However, we do not favour this approach, partly because it is not clear to us how to ensure that this phantom field does not appear at low energies, such that an observable runaway proceses is absent \cite{cline}. We do feel however that, inspite of their lack of utility for our purposes, that the two examples just covered, show that the question of whether or not scalar fields with wrong sign kinetic energies can be embedded into some well defined high energy theory, is sufficiently an open a question as to motivate the framework which we work in. 
\par
We now sketch how one migh embed our proposal in a higher derivative theory, for which a `wrong sign' kinetic energy for one of the fields is only a temporary feature.

\subsection{A higher derivative construction}

One could have distilled our discussion in section II, to the following observation -- all that we require for our proposal to work is, that there is some form of matter which satisfies the equation of state

\eq{eosp}{\rho = p,}

\noindent whilr $\dot{\rho} = 0$, and for which, by virtue of (\ref{00}),

\eq{nega}{\rho < 0}

\noindent holds at least temporarily. We realised this through the use of a scalar field which temporarily acquires the wrong sign kinetic energy. This is not to say that this is the only way in which such a situation might arise, but as the examples just furnished hoped to have illustrated, it is perhaps the best motivated. We take note that the `wrong sign' kinetic energy of a particular scalar field, might also be dynamically realised in a higher derivative setting, where our inflaton field is coupled to itself in the following manner:

\eq{coup}{S_{int} = \frac{-1}{2\lambda}\int d^{D+1}x \sqrt{-g}\bigl[f(\phi) (\nabla\phi)^2 + g(\phi)(\nabla\phi)^4 \bigr],}

\noindent where $\lambda$ is some appropriate dimensionful parameter and $f$ and $g$ are some as yet undetermined functions of $\phi$. Once could imagine a situation (which was studied in a different context in \cite{k}), where $f$ and $g$ are such, that one has for small enough values of $\dot\phi$, an effectively negative kinetic energy but for larger values of $\dot\phi$, the function $g$ is such that the Hamiltonian of the system remains bounded from below. Although seemingly similar, this approach is different from that taken in k-inflation \cite{k}, in that we only need these higher derivative terms in order to embed our inflaton in a well defined theory. The higher derivative terms themselvels do not play much of a roll in generating the inflationary phase, but might might play a roll in ending it.  
\par

Clearly the specifics of this, and some of the other approaches just discussed as a realization of our inflationary framework are involved enough, that we postpone their detailed investigation to a future report-- our goal in this report being to merely motivate $w = 1$ inflation as a viable alternative to potential-driven inflation. We now turn our attention towards understanding the cosmological consequences of our proposal.

\section{Stability and Small Parameters}

An immediate question that one might have, is whether or not the solution we propose is stable or not. So far, we have only presented $w = 1$ inflation in an abstract form, and a meaningful study of the question of the stability of our solution, and whether or not it corresponds to an attractor solution might depend on its precise realization. For instance, one may encounter differences in answering this question were $w = 1$ inflation realised through a propagating negative tension brane, or were it to be realised as being embedded in some well defined higher derivative theory, as there will be different degrees of freedom associated with either of these scenarios. Nevertheless, it is still meaningful to study the issue of the stability of our solution in the abstract form that it is presented in, as most of the scenarios we have proposed will likely be accurately modelled in some regime by our framework. Bearing this caveat in mind, we proceed by first re-examining the equations of motion:

\begin{eqnarray}
\label{1} \dot{H}_i &+& (\sum_j H_j)H_i = 0,\\
\label{2} \ddot{\phi} &+& (\sum_j H_j)\dot{\phi} = 0,\\
\label{3} -\dot{\phi}^2 &=& \frac{1}{8\pi G_D}\bigl[ (\sum_j H_j)^2 - \sum_j H_j^2\bigr].
\end{eqnarray} 

\noindent Recalling that the solution found earlier corresponded to $\dot{H}_i = \ddot{\phi} = 0$, $\sum_j H_j = 0$, we make the ansatz:

\begin{eqnarray}
\label{p1} H_i(t) &=& H_i^0 + h_i(t),\\
\label{p2} \dot{\phi}(t) &=& \dot{\phi}^0 + \varphi (t), 
\end{eqnarray}

\noindent where $h_i(t)$ and $\varphi (t)$ are small fluctuations around $H_i^0$ and $\dot{\phi}^0$, which are constants that satisfy (through the background equations of motion):

\begin{eqnarray}
\label{p3} \sum_j H^0_j &=& 0,\\
\label{p4} \frac{1}{8\pi G_D}\sum_j  H_j^{0^2} &=& \dot{\phi}^{0^2}. 
\end{eqnarray}

\noindent In what follows, although we have assumed that the fluctuations around our inflating solution are small, we do not neglect any higher order terms, as it turns out that we can solve the full non-linear differential equations analytically. We begin by substituting (\ref{p1}) and (\ref{p2}) into (\ref{1}) and (\ref{2}) respectively, where upon using (\ref{p3}) we find that the equations of motion become:

\begin{eqnarray}
\label{ss1} \dot{h}_i(t) + [\sum_j h_j(t)][H^0_i + h_i(t)] &=& 0,\\
\label{ss2} \dot{\varphi}(t) + [\sum_j h_j(t)][\dot{\phi}^0 + \varphi(t)] &=& 0.
\end{eqnarray}

\noindent We now sum (\ref{ss1}) over all indices, using (\ref{p3}) again to obtain the equation:

\eq{ome}{\dot\Omega + \Omega^2 = 0~~;~~ \Omega = \sum_j h_j(t),}

\noindent which has the solution:

\eq{omes}{\Omega = \frac{1}{t + c},}

\noindent where $c$ is some integration constant, which we evaluate at time $t=0$ as:

\eq{smp}{1/c = \sum_j h_j(0) := \epsilon.}

\noindent It is now a straightforward matter to solve for (\ref{ss1}) and (\ref{ss2}), and from there to obtain the full solutions of (\ref{1}) and (\ref{2}):

\begin{eqnarray}
\label{h1} H_i(t) &=& \frac{H_i(0)}{1 + \epsilon t}~~;~~ H_i(0)=h_i(0) + H^0_i\\
\label{ph1} \dot{\phi}(t) &=&  \frac{\dot{\phi}(0)}{1 + \epsilon t}~~;~~\dot{\phi}(0)=\varphi(0) + \dot{\phi}^0.
\end{eqnarray}   

\noindent Hence we see the significance of the small parameter $\epsilon$ introduced above: it determines whether or not we remain in a purely exponentially expanding/contracting phase, or whether or not we deviate away from it, and how quickly we do so. Realizing the nature of the equations (\ref{1}) and (\ref{2}) as first order differential equations, we see that it makes sense that $\epsilon = 0$ yields us an exponential inflating/deflating solution similar to the one we found earlier, except with different values for the Hubble factors and $\dot\phi$. This is because (\ref{p3}) and (\ref{p4}) define (for isotropic non-compact dimensions, and separately isotropic compact dimensions) a one parameter family of exponentially inflating/deflating solutions. That $\epsilon = 0$, by definition means that:

\eq{hjk}{\sum_j h_j(0) = 0,}

\noindent Hence the solution which we are perturbing towards, is simply another exactly exponential solution further along in parameter space. Thus it makes sense physically that deviations away from this solution are sensitive only to the combination $\epsilon$ that is non-vanishing. One can gain further insight into this small parameter, by further integrating (\ref{h1}), to obtain solutions for the metric factors:

\begin{eqnarray}
\label{es1}
a(t) &=& \lambda_a(1 + \epsilon t)^{H(0)/{\epsilon}},\\
\label{es2} b(t) &=& \lambda_b(1 + \epsilon t)^{\mathcal{H}(0)/{\epsilon}},
\end{eqnarray}    

\noindent where $a(t)$ and $b(t)$ are the scale factors for the non-compact, and the compact dimensions respectively, $\lambda_a$ and $\lambda_b$ are constants, and where $H(0)$ and $\mathcal{H}(0)$ are defined in (\ref{h1}). We see that because of the well known relation:

\eq{exp}{e^{x}= \lim_{n \to \infty}\bigl(1 + \frac{x}{n}\bigr)^n,}

\noindent that in the limit $\epsilon \to 0$, our analytic solutions interpolate smoothly to the solutions we found earlier:

\begin{eqnarray}
a(t) &=& \lambda_a e^{Ht},\\
b(t) &=& \lambda_b e^{\mathcal{H}t,}
\end{eqnarray}   

\noindent such that (\ref{vct}) is satisfied. Note that one tends to the above solutions independent of the sign of $\epsilon$, as one only has the combination $tH$ in the argument of the exponential. Hence we see that the smallness of the parameter $\epsilon$ governs (by virtue of (\ref{exp})) how close to exactly exponential our solution is. One can make this statement as precise as we wish in a functional analytic sense due to well known properties of the relation (\ref{exp}). We should also note that since the solutions (\ref{es1}) and (\ref{es2}) are exact, that even for large values of $\epsilon$, one still has an inflating/deflating solution provided that $H_0$ and $\mathcal{H}_0$ are large enough initially, except that now we will in general have power law or pole law inflation. However, we are most interested in the case where $\epsilon$ is small, and in this case, one can extract a timescale from it, which will govern for how long we have a period of de Sitter expansion (contraction) of the non-compact (compact) dimensions:

\eq{timesc}{\tau = \frac{1}{\epsilon}.}

\noindent The smaller $\epsilon$ is, the longer we exist in this de Sitter phase. We should take note here, that since our inflaton is describing some temporarily unstable configuration, we will have in genereal, another timescale to contend with, the lifetime of this instability. It will be the shorter of these two timescales that will determine how much we inflate. 

\section{Kinetic De Sitter Inflation -- So What?}

We first note that in proposing this inflationary framework, we have intrinsically tied up the fact that we have had a period of inflation with the notion that extra dimensions, if they exist, must be very small. That is, we have uncovered a very natural dynamical compactification mechanism. Since string theory wants to set up a hierarchy between compact and non-compact dimensions, we find that this is certainly one  way of doing it (see \cite{BV}\cite{randall}\cite{mazumdar}\cite{durrer} for other such ideas). In fact, the amount of e-foldings in this type of inflation is either dertermined by the lifetime of the scalar field we use as our inflaton, or the timescale for which it has a `wrong sign' kinetic energy, or is determined by $\epsilon$ (\ref{smp}), or by the relation uncovered in section II:

\eq{vc}{dH + n\mathcal H = 0.}

\noindent That is, if we want $Ht \sim 60$ then $\mathcal H t \sim -d/n 60$, which for $d = 3$, $n = 6$ implies $\mathcal H = -20$. So the compact dimensions must shrink by 20 e-foldings. Of course, the naturalness of this requirement depends on the exact setting in which our model of inflation occurs. For instance, it could be that the universe has some sort of a standard big bang, and inflation sets in at a later epoch, as it must if it is to solve the monopole problem. In this case, Inflation sets in and sets the scene for the `hot big bang' that proceeds after inflation ends, except now any extra dimensions that may have been around are unobservably tiny (we discuss in the next section how a string theoretical realization of this framework can naturally keep these extra dimensions stabilized at the string scale). 

\par

To recall the discussion in the introduction to this report, we have now obtained a inflationary epoch without the use of potentials, and as such, have freed up our search for the solution to the cosmological constant problem somewhat. As a consequence of this, we might have also avoided, or at least ameliorated some of the fine tuning one has to do (depending of course on the specific realization of our mechanism), in that there is now, no obvious place to structurally tune the model, except in the possible lifetime of `wrong sign' phase. The parameter $\epsilon$, is the only apparent quantity we have to tune, which in fact may be constrained by the specific realization of $w = 1$ inflation. In fact, if one were to take the lifetime of the wrong sign phase to be longer than what it takes for the extra dimensions to shrink to some fundamental scale at which new physics is bound to set in and end this type of inflation (see next section), then the amount of e-foldings that occur for the non compact dimensions is completely determined by the relation (\ref{vc}), and the initial size of the compact dimensions, as discussed above. However, a potential concern of having a purely de Sitter phase of inflation, is that aside from its conceptual appeal, it may not make much in the way of testable predictions that differ from most models of potential-driven inflation. That is, how do we know that we do not live in a completely `vanilla universe'\cite{easther}? The answer to this, shall concern us next.

\section{Graceful Exit and Reheating}

We mention two avenues for a graceful exit from this type of inflation. The first makes use of the finite lifetime of the 'wrong sign' phase of the inflaton. Clearly, inflation will end once the sign of the kinetic energy returns to being positive definite (or the inflaton decays). This is a likely place where further parameters will enter into the calculation. In order to be specific, consider the example of a negative tension brane propagating through a compact orbifold. The maximum time such a brane could exist, is given by:

\eq{tau}{\tau = \frac{\pi}{\dot{\theta}_0},}

\noindent where we recall that an orbifold is defined by the interval $-\pi \leq \theta \leq \pi$, which the identification $\theta \leftrightarrow -\theta$, with $\theta$ the appropriate angular variable, and $\dot{\theta}_0$ some initial angular velocity. How large this lifetime is will govern how much inflation we get. Note that this is strictly an upper bound, but is nevertheless demonstrative of how one can determine such a lifetime.

\par

Another manner in which inflation can end is through new effects that appear once the extra dimensions reach the minimal length of the model in question. For instance, in the context of heterotic or bosonic string theory, it has been shown in\cite{sp1}\cite{sp2}\cite{sp3}\cite{scott} that new massless states that appear at the string scale are likely to condense and stabilize any number of extra dimensions at the string scale. The fact that these degrees of freedom are massless, implies that, provided we stay at weak string coupling, the production of these modes in reheating should not be a problem. In \cite{sp2}, it was shown that a period of Friedmann-Robertson-Walker (FRW) expansion of the non-compact dimensions then ensues. If this type of inflation were to find a string theoretical realization, then moduli stabilization  inherent to massless string cosmology would provide us with a natural graceful exit and reheating mechanism, along with possible observational consequences for the CMBR \cite{ts}. 

\par

However, we do not wish to tie down the discussion to a particular mechanism, as we wish to point out in general, what if any predictions this framework of inflation might make that could be potentially testable, and contrasted against potential-driven inflation. Clearly, if the scalar field that we use as our inflaton is describing some unstable degree of freedom, then the details of the demise of this degree of freedom will have definite consequences in terms of reheating that ensues when inflation ends. Hence although it was our goal in this paper to demonstrate how a period of kinetic inflation can sidestep some of the conceptual problems that potential-driven inflation encounters, it is in the specifics of particular realization of this framework that one is going to encounter mechanisms which offer testable predictions that can be contrasted against the predictions of potential-driven inflation. Although this will certainly occupy us in a future report, our hope is that the results contained in this paper will inspire others to look for inflation in places where they might not have otherwise thought it possible.     

\section{Conclusions and Outlook}

In this paper, we have presented a form of kinetic inflation, that is driven by a scalar field which experiences a short lived phase in which it acquires a `wrong sign' kinetic energy term. One of the advantages of this approach is that one might not have to do the same amount of fine tuning that one has to do for the usual potential-driven inflation, in addition to providing a natural dynamical compactification mechanism for our universe. The fact that this mechanism avoids using potentials also implies that one does not have to worry that the eventual solution of the cosmological constant problem will invalidate our inflationary mechanism. However, the main advantage of this approach, from a phenomenologist's perspective, is that it offers a new window in which to search for inflation in string theory, as it is a theory which admits a plethora of (in the absence of any other mechanism) massless moduli fields, that could behave in the manner required by this framework. At the very least, we hope to have indicated an interesting use of scalar fields with temporarily `wrong sign' kinetic energies, and have added to the list of novel ways in which one can obtain inflation.

\section{acknowledgements}

We wish to thank Cliff Burgess, Jim Cline, Joshua Elliot and Guy Moore for valuable input, criticisms and discussions during the early stages of this work. We also wish to thank Thorsten Battefeld, Mark Trodden, and especially Damien Easson for suggestions and scurtiny of this manuscript. Foremost, we wish to thank Robert Brandenberger for continued support, inspiration, patience and not least, many useful discussions. This work is supported by an NSERC discovery grant at McGill University as well as funds from McGill University, which have been made possible by the generous support of the government, and the people of Canada, to whom this author is indebted.

\end{document}